\documentclass[aps,draft]{revtex4}
\begin{document}
\draft
\title{Finite-temperature correlations in the one-dimensional\\
          trapped and untrapped Bose gases}
\author{N.M. Bogoliubov\footnote{E-mail: \it{bogoliub@pdmi.ras.ru}}}
\address{Steklov Institute of Mathematics at St.Petersburg, Fontanka 27,\\
St.Petersburg 191011, Russia}
\author{R.K. Bullough\footnote{E-mail: \it{robin.bullough@umist.ac.uk}}}
\address{Department of Mathematics,\\
The University of Manchester Institute\\
of Science and Technology, P.O. Box 88, Manchester M60 1QD, United Kingdom}
\author{C. Malyshev\footnote{E-mail: \it{malyshev@pdmi.ras.ru}}}
\address{Steklov Institute of Mathematics at St.Petersburg, Fontanka 27,\\
St.Petersburg 191011, Russia}
\author{J. Timonen\footnote{E-mail: \it{timonen@phys.jyu.fi}}}
\address{Department of Physics, University of Jyv\"askyl\"a, P.O. Box 35,\\
FIN-40351, Finland}


\begin{abstract}

We calculate the dynamic single-particle and many-particle correlation
functions at non-zero temperature in one-dimensional trapped repulsive 
Bose gases.
The decay for increasing distance between the points of these correlation
functions is governed by a scaling exponent that has a universal
expression in terms of observed quantities. This expression is valid in
the weak-interaction Gross-Pitaevskii as well as in the strong-interaction
Girardeau-Tonks limit, but the observed quantities involved depend on the
interaction strength. The confining trap introduces a weak center-of-mass
dependence in the scaling exponent. We also conjecture results for the
density-density correlation function.

\end{abstract}


\maketitle

\section{Introduction}

Recent advances in experimental techniques have made it possible to study
Bose gases in what are effectively two-dimensional and even one-dimensional 
(1D) traps \cite{g}, \cite{sh}, \cite{grein}, \cite{bongs}. The confinement 
of the three-dimensional (3D) gas achieved by making the level spacing of 
the confining potential in one or two dimensions larger than the energy of
the individual atoms. Because of the growing interest in coherent
matter-wave interferometry and atom lasers, these recent developments
have also revived  theoretical interest in the properties of these 
effectively 1D Bose systems. It is clearly important to understand in 
more detail the effects of confinement on, {\it e.g.}, their ground-state 
and thermal properties including one-particle as well as many-particle 
correlations. It is well established \cite{pethic} that a good 
theoretical framework for ultracold metal vapours is given by bosons 
with (in the case considered here) repulsive delta-function 
interactions. We can thus describe the 1D system with a Hamiltonian
\begin{equation}
\hat H=\int \left\{ \hat \psi ^{\dagger }\left( x\right) \left( \frac{-\hbar
^2}{2m}\frac{\partial ^2}{\partial x^2}-\mu +V\left( x\right) \right)
\hat \psi \left( x\right) +\frac g2 \hat \psi ^{\dagger }\left( x\right)
\hat\psi^{\dagger }\left( x\right) \hat \psi \left( x\right) \hat \psi
\left(x\right) \right\} dx,
\label{ham}
\end{equation}
where $\hat \psi (x)$ is a Bose field operator with its adjoint
$\hat \psi^{\dagger }(x)$ and commutation relations
$[\hat \psi (x),\hat \psi ^{\dagger }(x^{\prime })]
    =\delta (x-x^{\prime })$, {\it etc.},
$\mu $ is the chemical potential, $g>0$
is the interaction strength (the coupling constant), and
$V(x)\equiv \frac m2\Omega ^2x^2$ is the harmonic trap potential.
For $V(x)\equiv 0$ this system has been a prototype for exactly solvable
models in 1D \cite{gir}, \cite{liebl}, \cite{yang}, \cite{kor},
\cite{bultim}, and many of its properties have already
been worked out in great detail.

In real physical systems the transition from 3D to 1D behaviour has an
interesting aspect in that the resulting effective density of atoms in 1D
can be either `high' or `low' depending on the parameters of the system.
It is a feature of 1D Bose systems that high density means weak
interactions and low density strong interactions between the particles.
This property opens up the possibility of realising experimentally the
'strong-coupling' limit \cite{gan} in which repulsive bosons display
fermionic properties \cite{gir}, \cite{liebl}. This is the so-called
Girardeau-Tonks regime. In this 'true' 1D regime in the sense of 
\cite{lieb2}, the Bose-Einstein condensation in the sense of 
\cite{lieb2} is not realized, and the system is not coherent.

In contrast, in the limit of high effective 1D densities of atoms,
{\it i.e.}, in the Gross-Pitaevskii, or the weak-interaction
regime, short-range coherence builds up between the atoms, and the
system forms at low enough temperatures a {\it quasi} {\it
condensate} \cite{pop}, \cite{homa}, \cite{pet}, \cite{bogoli}, 
\cite{det}, \cite{shv}, \cite{gerb}, \cite{gerb2}.
True long-range order is in this regime destroyed by phase
fluctuations, while density fluctuations are suppressed. It is
thus evident that the local correlations between the atoms are
different in these two regimes such that the `local' two-point
correlation function $\Gamma (x_1,x_1)$ tends to zero in the
Girardeau-Tonks regime, and to unity in the Gross--Pitaevskii
regime with quasi condensate \cite{kher}. At high temperatures or
in the very small coupling-constant limit it obviously becomes
two, the value for uncorrelated bosons. Of course we are not
limited to two-point correlations: thus, {\it e.g.}, the local three-point
correlation function $\Gamma (x_1,x_1,x_1)$ is important because
it is related to three-body recombination, an inelastic decay
process \cite{gan}. Notice that in defining the full non-local
correlation functions we need as many space variables
$x_1,x_2,...,x_n$ as are being correlated because in the presence
of a confining potential there is no longer translational
invariance. Moreover at finite temperatures it becomes natural to
correlate also time-like variables $\tau$ for which the fields
evolve in Matsubara representation $\hat \psi(x,\tau )=e^{-\tau
\hat H}\hat \psi (x)e^{\tau \hat H}$ with $\hat H$ given by
(\ref{ham}) and likewise for $\hat\psi ^{\dagger}(x,\tau)$ \cite{lif}.
Thus the two-point (single-particle) {\it thermal} correlation 
function which will be studied in this paper is given by
\begin{equation}
\Gamma (x_1,\tau _1;x_2,\tau _2)=\left\langle
T_\tau \hat \psi ^{\dagger}(x_1,\tau _1)\hat \psi (x_2,\tau _2)
\right\rangle ,
\label{tag2}
\end{equation}
and the $n$-point correlation function correspondingly.
In these expressions $T_\tau $ is a `time-ordering' operator
which places the field operators from right to left in order
of increasing $\tau$ \cite{lif}. Note that these expressions
like Eq. (\ref{tag2}) are not translationally invariant in $x$
but they are translationally invariant in  $\tau$. Variables
$\tau$ lie in $0\leq \tau \leq \beta ,\beta \equiv(k_BT)^{-1}$
and $T$ is temperature, and there is (for bosons) periodicity 
in $\tau$  period $\beta$.

Despite the previous results for one or more
space dimensions \cite{pet}, \cite{bogoli}, \cite{gl}, \cite{khaw},
\cite{an}, \cite{petr}, \cite{lux}, \cite{dunj} the decay of the 
single-particle and especially many-particle correlations in the 
presence of a confining potential have not been fully analysed. We 
address this problem for one space dimension in this paper, and
point out that the way these correlations decay for increasing 
$|x_1-x_2|$ is governed by a {\it universal} scaling exponent 
that can be expressed in terms of observed quantities, and, at finite 
temperature, applies in both of the fundamental regimes described 
in \cite{lieb2}. This scaling exponent is analogous to the one that
appears in the power-law decay of correlations in the homogeneous system
of 1D bosons at zero temperature. In the presence of a confining potential,
this exponent becomes a function of the center-of-mass coordinate, but 
this dependence can be expected to be small. This result is valid in the
weak-interaction Gross-Pitaevskii as well as the strong-interaction
Girardeau-Tonks regime and thus straddles both of the
two regimes distinguished in \cite{lieb2}.  
We also show how asymptotically all multi-particle correlations
can be expressed in terms of the two-point correlations, and indicate
how the results can be extended to density-density correlations. 
As all of these properties can be solved exactly by Bethe Ansatz 
techniques in the translationally invariant case of no confinement, 
this case provides a good point of reference, and
we begin by reviewing it in the next section.

Finally, the existence of the experiments like \cite{g}, \cite{sh},
\cite{grein}, \cite{bongs}, providing essentially one dimensional 
systems, means that it could become possible to check connected one 
dimensional theory in great detail at finite temperatures. A purpose 
of this paper is thus to point out that the methods employed here can
be used to provide such a connected theory.

\section{Summary of exact results by the Bethe Ansatz method}

Let us consider the system described by the Hamiltonian (\ref{ham}). In the
absence of the confining potential namely by setting $V(x)\equiv 0$ the
Hamiltonian (\ref{ham}) becomes exactly solvable for its eigenstates and
eigenvalues by the Bethe Ansatz method \cite{kor}. For a gas of $\ N$ bosons
on a ring of circumference $L,$ namely for periodic boundary conditions of
period $L$ in the one dimensional space $x,$ eigenfunctions of Hamiltonian
(\ref{ham}) are \cite{liebl},\cite{kor}:
\begin{equation}
|\Phi _N(\lambda _1,...,\lambda _N)\rangle =\frac 1{\sqrt{N!}}\int
dx_1...dx_Nf_N(x_1,...,x_N\mid \lambda _1,...,\lambda _N)\hat \psi^{\dagger
}(x_1)...\hat \psi ^{\dagger }(x_N)\mid 0\rangle ,  \label{cefun}
\end{equation}
and these depend on $N$ real parameters $\lambda _1,...,\lambda _N$ \
satisfying the so-called Bethe equations
\begin{equation}
e^{i\lambda _jL}=-\prod_{k=1}^N\frac{\lambda _j-\lambda _k
+i(\tilde g/2)}{\lambda _j-\lambda _k-i(\tilde g/2)}\,;\,\,\,j=1,...,N,  
\label{cbethe}
\end{equation}
where $\tilde g=(2m/\hbar ^2)g$ is the renormalised coupling constant and
\begin{equation}
f_N=C_N\sum_{{\cal P}}\exp \left\{i\sum_{n=1}^Nx_n\lambda _{{\cal P}%
_n}\right\}\prod_{1\leq k<j\leq N}\left[ 1-\frac{\tilde g}2\frac{i\epsilon
(x_j-x_k)}{(\lambda _{{\cal P}_j}-\lambda _{{\cal P}_k})}\right] .
\label{cwfun}
\end{equation}
Here the sum over all the permutations ${\cal P}$ of the numbers $1,2,...,N$
is taken, $\epsilon (x)=sign(x)$ is the sign function, and the factor $C_N$
is equal to
\[
C_N=\frac{\prod\limits_{j>k}(\lambda _j-
\lambda _k)}{\sqrt{N!\prod\limits_{j>k}
\bigl[(\lambda_j-\lambda _k)^2+(\tilde g/2)^2\bigr]}}.
\]
The Fock vacuum $|0\rangle $ is defined by the condition
$\hat\psi (x)|0\rangle=0.$ The corresponding eigenenergies are
\begin{equation}
E_N=\frac{\hbar ^2}{2m}\sum_{j=1}^N\lambda _j^2.  \label{cen}
\end{equation}

The solutions $\lambda _j$ of the Bethe equations, Eq. (\ref{cbethe}),
multiplied by $\hbar $ have a natural interpretation as the momenta of the
particles each of mass $m$. The function $f_N$ is a symmetric function of
the variables $x_j$ and a continuous function of each $x_j.$ One can also
see that $f_N$ is an antisymmetric function of $\lambda _j.$ Hence, $f_N=0$
if $\lambda _j=\lambda _k$, $j\neq k$, in the whole coordinate space
$-\infty <x_j<\infty $ for each $j$. This property is the basis of the Pauli
principle for one-dimensional interacting bosons. The wave functions (\ref
{cwfun}) form a complete and orthonormal set in configuration space and thus
the eigenfunctions (\ref{cefun}) are orthogonal for the different sets of
the solutions of the Bethe equations. At zero temperature ($T=0$) the ground
state of the system is defined by the solutions lying within the interval 
$-\lambda _F\leq \lambda _j\leq \lambda _F$ and these form a ''Fermi sphere''
in the momentum space. At an arbitrary value of the temperature ($T\neq 0)$
the solutions are distributed along the real axis:\ $-\infty <\lambda
_j<\infty $. In the thermodynamic limit in which the number of particles $N$
and the circumference of the ring $L$ both go to infinity, $N,L\rightarrow
\infty ,$ at a fixed density $\rho =\frac NL,$ the distribution function of
the solutions of the Bethe equations is defined by the solutions of the
Lieb-Liniger equations for $T=0$ \cite{liebl} and by the solutions of the
Yang-Yang equations for $T>0$ \cite{yang}. The free energy density ${\cal E}$
of the gas is ${\cal E}=\lim_{N,L\rightarrow \infty }E_N/L.$ The ground
state with a particle density $\rho >0$ corresponds to positive values of
the chemical potential in this exactly solvable case.

In \cite{kor} it is proved that the correlation function ($\tau _1<\tau _2$)
\begin{equation}
\langle \hat \psi ^{\dagger }(x_1,\tau _1)\hat 
    \psi (x_2,\tau _2)\rangle \equiv
\frac{{\rm tr}\left( 
e^{-\beta \hat H}\hat \psi^{\dagger }(x_1,\tau _1)\hat \psi
(x_2,\tau _2)\right) }{{\rm tr}\left( e^{-\beta \hat H}\right) }  
\label{ccf}
\end{equation}
may be expressed in the form
\begin{equation}
\langle \hat \psi ^{\dagger }(x_1,\tau _1)\hat \psi (x_2,\tau _2)\rangle =
\frac{\langle \Omega _T\mid \hat \psi ^{\dagger }(x_1,\tau _1)\hat \psi
(x_2,\tau _2)\mid \Omega _T\rangle }{\langle \Omega _T\mid \Omega _T
\rangle },
\label{ccfte}
\end{equation}
where $\mid \Omega _T\rangle $ is any of the eigenfunctions (\ref{cefun})
contributing in the thermodynamic limit to the state of thermal equilibrium.
It depends on temperature through the distribution function of the solutions
of the Bethe equations. For small nonzero temperatures and long distances
$\mid x_1-x_2\mid \gg l_c$ (here $l_c$ is a healing length
$l_c=\hbar /\sqrt{m\rho g}$) the asymptotics of the correlator
(\ref{ccf}) is
\begin{equation}
\langle \hat \psi ^{\dagger }(x_1,\tau _1)\hat \psi (x_2,\tau _2)\rangle
\simeq \frac \rho {\left| \sinh \frac \pi {\hbar \beta v}\left( \left|
x_1-x_2\right| +i\hbar v(\tau _1-\tau _2)\right) \right| ^{1/\theta }},
\label{tcor}
\end{equation}
in which $\theta $ is the scaling exponent
\begin{equation}
\theta =\frac{2\pi \hbar \rho }{mv},  \label{isv}
\end{equation}
and $v$ is the sound velocity, $v^2=(\rho/m) \frac{\partial ^2{\cal E}}
{\partial \rho ^2}$. In this form the scaling exponent depends only on
observable quantities, namely the density $\rho ,$ the sound velocity
$v$, and the atomic mass $m$. The asymptotic result Eq. (\ref{tcor})
with the scaling exponent (\ref{isv}) is valid for arbitrary values of
the coupling constant $0<g\leq \infty $. For the so-called
Girardeau-Tonks gas ($g=\infty$) \cite{gir},\cite{lieb2},\cite{olsh} 
Appendix B shows that
\begin{equation}
\theta =2  \label{tg}
\end{equation}
exactly as is well known \cite{kor}. For small coupling constants $g$ we may
express (\ref{isv}) in the form \cite{bogol}
\begin{equation}
\theta =2\pi \hbar \sqrt{\frac \rho {mg}}.  \label{sgexp}
\end{equation}
One may also express $\theta $ (12) through the healing length $l_c:\,\theta
=2\pi \rho l_c$. In Appendix B we also give a simple derivation of these
expressions (11), (12).

Equation (\ref{tcor}) means that the correlator (\ref{ccf}) decays
exponentially at long distances $|x_1-x_2|\gg l_c$ ,
\begin{equation}
\langle \hat \psi ^{\dagger }(x_1,\tau _1)\hat \psi (x_2,\tau _2)\rangle
\sim \rho \exp \left(-\frac \pi {\hbar \beta v\theta }\left| x_1-x_2
\right|\right),
\label{itcor}
\end{equation}
with the correlation length
\begin{equation}
\xi =\frac{2\hbar ^2\beta \rho }m=\frac{\hbar \beta v}\pi \theta .
\label{icorl}
\end{equation}
The result Eq. (\ref{tcor}) also allows us to obtain the asymptotic
behaviour of the correlation function at zero temperature. In the zero
temperature limit ($\beta \rightarrow \infty $) the correlation function
(\ref{tcor}) transforms into the correlation function for vacuum 
fluctuations and has a power law decay
\begin{equation}
\langle \hat \psi ^{\dagger }(x_1,\tau _1)\hat \psi (x_2,\tau _2)\rangle
_{T=0}\simeq \frac \rho {\Bigl| | x_1-x_2 | +i\hbar v(\tau _1-\tau
_2)\Bigr| ^{1/\theta }}.  \label{iztcf}
\end{equation}

The distinction in the behaviour of the correlators at zero and finite
temperatures \ may be explained by the formation of the Fermi sphere at 
zero temperature \cite{kor}. However this ground state of bosons cannot be
considered as a true condensate since $\langle \hat \psi ^{\dagger}
(x_1,\tau _1)\hat \psi (x_2,\tau _2)\rangle $ vanishes when
$|x_1-x_2|\rightarrow \infty $ so that there is no long-range order in any
general sense. As correlations decay algebraically, {\it i.e.}, 
by a power law the system is scale invariant and can be thought 
to be at a critical point with $\theta $ now a critical exponent.

\section{The functional integral approach}

\subsection{The partition function}

We now evaluate strictly comparable results to those summarized in Section
II by using methods of functional integration. We also obtain results which
are more general than those of Section II in that the effects of the
confining potential $V(x)=\frac m2\Omega ^2x^2$ are taken into account.

The partition function $Z$ for a one-dimensional gas with repulsive
$\delta$-function interactions of strength $g>0$ is given as a
functional integral by \cite{pop},\cite{st}
\begin{equation}
Z=\allowbreak \,\int e^{S[\psi ,\bar \psi ]}{\cal D}\psi {\cal D}
\bar\psi .
\label{part}
\end{equation}
The {\it classical} action $S$ of this system is equal to
\begin{equation}
S[\psi ,\bar \psi ]=\int_0^\beta d\tau \int dx\left( \bar \psi (x,\tau )
\hat K\psi (x,\tau )-\frac g2\bar \psi (x,\tau )
                \bar \psi (x,\tau )\psi (x,\tau)\psi (x,\tau )\right) .  
\label{act}
\end{equation}
The differential operator in this action $S$ is $\hat K=\frac \partial 
{\partial \tau }-{\cal H}$, while the one particle Hamiltonian ${\cal H}$
includes the trap potential $V(x)=\frac m2\Omega ^2x^2$ and the chemical
potential $\mu $ is also contained in ${\cal H}$ which is
${\cal H}\equiv-\frac{\hbar ^2}{2m}\frac{\partial ^2}{\partial x^2}+V(x)-\mu$. 
We have introduced two $c$-number
complex valued fields $\psi (x,\tau ),\bar \psi (x,\tau )$ and, consistent
with the existence of the potential $V(x),$ the boundary conditions for
these fields are chosen to be vanishing at infinity for $x$ (in the sense of
quadratic integrability) and periodic, period $\beta =(k_BT)^{-1}$, for $%
\tau $. These new $c$-number complex valued fields $\psi ,\bar \psi $ are
two independent fields, and they are introduced in a formal correspondence
with the operator fields $\hat \psi (x,\tau )$, $\hat \psi ^{\dagger}
(x,\tau )$ \cite{pop}. In (\ref{part}), ${\cal D}\psi {\cal D}\bar \psi $ is
the integration measure.

For small enough temperatures we can expect to put $\psi (x,\tau )=\psi
_o(x,\tau )+\psi _e(x,\tau )$ and likewise for $\bar \psi $. The condensate
variable $\psi _o(x,\tau )$ describes the condensate field and $\psi
_e(x,\tau )$ describes highly excited thermal particles. Also we can expect
to write the field variable $\psi _o(x,\tau )=\psi _o(x)+\xi (x,\tau ),$
where the field $\psi _o(x)$ describes the ground state of the model at zero
temperature (for the homogeneous case this field describes the particles in
the Fermi sphere), while the field $\xi (x,\tau )$ describes the low lying
excited particles (the phonon-like excitations in the vicinity of the Fermi
momenta in the homogeneous case). We shall also require that the fields
$\psi _o,\psi _e$ are orthogonal in the sense: $\int dx\psi _o(x,\tau )\bar
\psi _e(x,\tau )$ $=$ $\int dx\bar  \psi _o(x,\tau )\psi _e(x,\tau )=0$ .
Since by construction $\psi _o$ and $\psi _e$ are orthogonal, and 
likewise for $\bar \psi_o$ and $\bar \psi _e$, we can write the 
integration measure as ${\cal D}\psi {\cal D}\bar \psi 
={\cal D}\psi _o{\cal D}\bar \psi _o{\cal D}\psi _e {\cal D}\bar \psi _e$.

We shall only consider terms in $S$ up to quadratic (bilinear) in 
$\psi _e$,$\bar \psi_e.$ This means that we are making an 
approximation in which the non-condensed particles do not interact 
with each other. By doing this we can then actually integrate out 
the thermal fluctuations and obtain the effective action functional
\begin{equation}
S_{eff}[\psi _o,\bar \psi _o]=\allowbreak \,\ln \int e^{S[\psi ,\bar \psi ]}
{\cal D}\psi _e{\cal D}\bar \psi _e,  \label{effact}
\end{equation}
depending only on the $\psi _o,\bar \psi _o$ variables. Then,
\begin{equation}
Z=\allowbreak \,\int e^{S_{eff}[\psi _o,\bar \psi _o]}{\cal D}\psi _o{\cal D}
\bar \psi _o.  \label{effpart}
\end{equation}

The effective action to lowest order in $g$ is given by 
(see Eq. (A15) Appendix A):
\begin{eqnarray}
&&\ S_{eff}[\psi _o,\bar \psi _o]=-\beta F_{nc}(\mu )  \nonumber \\
\ &+&\int_0^\beta d\tau \int dx\left\{ \bar \psi _o(x,\tau )
\left(\frac{\hbar ^2}{2m}\frac{\partial ^2}{\partial x^2}+
\Lambda -V(x)\right)\psi _o(x,\tau )-\frac g2 \bar \psi _o(x,\tau )
            \bar \psi _o(x,\tau )\psi _o(x,\tau )\psi _o(x,\tau)\right\},  
\label{seff}
\end{eqnarray}
where $\Lambda =\mu -2g\rho _{nc}(0)$ is a renormalized chemical potential
and $F_{nc}(\mu )$ is the free energy of the non-ideal gas of 
thermal particles (see Appendix A). 
At this point it is reasonable to pass to new variables, namely the density
$\rho (x,\tau )$ and the phase $\varphi (x,\tau )$ of the field
$\psi_o(x,\tau )$. These variables are defined through, and compare 
\cite{pop},
\[
\psi _o(x,\tau )=\sqrt{\rho (x,\tau )}e^{i\varphi (x,\tau )},\qquad \bar \psi
_o(x,\tau )=\sqrt{\rho (x,\tau )}e^{-i\varphi (x,\tau )},
\]
where $\rho $ and $\varphi $ are two independent real fields. 
Now the integration measure 
${\cal D}\bar \psi_o(x,\tau ){\cal D}\psi _o(x,\tau )$ is changed to
${\cal D}\rho (x,\tau ){\cal D}\varphi (x,\tau)$. In these new 
variables the effective action (\ref{seff}) will be equal to
\begin{eqnarray}
S_{eff}[\rho ,\varphi ] &=&-\beta F_{nc}(\mu )+i\int_0^\beta d\tau \int
dx\left\{ \rho \partial _\tau \varphi +\frac{\hbar ^2}{2m}\partial _x\left(
\rho \partial _x\varphi \right) \right\}  \nonumber \\
&&\ +\int_0^\beta d\tau \int dx\left\{ \frac{\hbar ^2}{2m}\left( \sqrt{\rho }%
\partial _x^2\sqrt{\rho }-\rho (\partial _x\varphi )^2\right) +\left(
\Lambda -V\right) \rho -\frac g2\rho ^2\right\} .  \label{sseff}
\end{eqnarray}
Here and below we denote the first order partial derivatives over $\tau $
and $x$ as $\partial _\tau $ and $\partial _x$, respectively, 
and the second order ones as $\partial _\tau ^2$ and $\partial _x^2$.

We consider the stationary phase approximation to Eq. (\ref{effpart}). The
appropriate extremum condition $\delta (S_{eff}[\rho ,\varphi ])=0$ for the
effective action (\ref{sseff}) has the form of two thermal,
$\tau$-dependent, Gross-Pitaevskii equations, namely
\begin{eqnarray}
 i\partial _\tau \varphi +\frac{\hbar ^2}{2m}\left( \frac 1{\sqrt{\rho }}
\partial _x^2\sqrt{\rho }-\left( \partial _x\varphi \right) ^2\right)
+\left( \Lambda -V\right) -g\rho &=& 0, \nonumber  \\
-i\partial _\tau \rho +\frac{\hbar ^2}m\partial _x\left( \rho \partial_x\varphi \right)
&=& 0.  \label{gp}
\end{eqnarray}
The substitution of the solutions $\rho _0,\varphi _0$ of these equations
into the effective action gives
\begin{equation}
S_{eff}[\rho _0,\varphi _0]=-\beta F_{nc}(\mu )+\frac g2\int_0^\beta d\tau
\int dx\rho _0^2.  \label{fe1}
\end{equation}
Here $F_{nc}(\mu )$ is the free energy of the non-ideal gas of
thermal particles. 
The total free energy of the system is thus \cite{bogoli}
$F(\mu )=-\beta^{-1}S_{eff}[\rho _0,\varphi _0]$.

At the Thomas-Fermi approximation \cite{pethic}, which is valid at low
enough temperatures the kinetic energy term
$(\partial _x^2\sqrt{\rho })/\sqrt{\rho }$ (so-called quantum pressure) in the
first equation
of Eqs. (\ref{gp}) may be omitted. The equilibrium solution relative to the
stationary ground state with $\partial _\tau \rho =0=\partial _\tau \varphi$
is obtained by setting the velocity field ${\bf v}=m^{-1}\partial _x\varphi$
equal to zero, ${\bf v}=0.$ The density profile is then given by,
and compare with Eq. (A7) where $\mu$ is now replaced by $\Lambda$, 
\begin{equation}
\rho _{TF}(x)\equiv\frac 
\Lambda g\tilde \rho _{TF}(x)=\frac \Lambda g \Bigl( 1-%
\frac{x^2}{R_c^2}\Bigr) \Theta \Bigl(1-\frac{x^2}{R_c^2}\Bigr),  
\label{tf}
\end{equation}
in which $\Theta $ is the Heaviside step function, and (recalling that
$V(x)=\frac m2\Omega ^2x^2$) $R_c$ is the radius of the ground state
at zero temperature, $R_c^2=\frac{2\Lambda }{m\Omega ^2}$.

Following the original decomposition $\psi _o(x,\tau )=\psi _o(x)+\xi
(x,\tau )$ we suppose that the thermal fluctuations in the vicinity of the
stationary state are small so that we may split the density profile as
\begin{equation}
\rho _0(x,\tau )=\rho _{TF}(x)+\pi _0(x,\tau ).  \label{thro}
\end{equation}
The Gross-Pitaevskii equations (\ref{gp}) linearized around the equilibrium
solution $(\rho _0=\rho _{TF},\varphi =const)$ will then take the form
\begin{eqnarray}
\ i\partial _\tau \varphi _0-g\pi _0+\frac{\hbar ^2}{4m\rho _{TF}}\partial
_x^2\pi _0 &=&0,  \nonumber \\
i\partial _\tau \pi _0-\frac{\hbar ^2}m\partial _x\left( \rho _{TF}\partial
_x\varphi _0\right) &=&0.  \label{string}
\end{eqnarray}
Eliminating $\varphi _0$ and dropping the term proportional to $\hbar ^4$,
we arrive at the thermal Stringari equation \cite{str} which is
\begin{equation}
\frac 1{\hbar ^2v^2}\partial _\tau ^2\pi _0+\partial _x\left( \Bigl( 1-
\frac{x^2}{R_c^2}\Bigr) \partial _x\pi _0\right) =0,  
\label{str}
\end{equation}
where $v$ is the sound velocity at the center of the trap,
\begin{equation}
v^2=\frac{\rho _{TF}(0)g}m=\frac \Lambda m.  \label{sv}
\end{equation}

The substitution $\pi _o=e^{i\omega \tau }u(x)$ transforms Eq. (\ref{str})
 into a Legendre-like equation
\begin{equation}
-\frac{\omega ^2}{\hbar ^2v^2}u(x)+\frac d{dx}\left( \Bigl( 1-\frac{x^2}{%
R_c^2}\Bigr) \frac d{dx}u(x)\right) =0.  \label{leg1}
\end{equation}
Since the Thomas-Fermi profile (\ref{tf}) differs from zero only within
$|x|<R_c$, we consider this equation for the fluctuating part $u(x)$ also at
$|x|<R_c$. After the analytical continuation $\omega \rightarrow iE$, the
equation (\ref{leg1}) possesses polynomial solutions, namely the
Legendre polynomials $P_n( x/R_c),$ if and only if
\[
\Bigl( \frac{R_c}{\hbar v}\Bigr) ^2E^2\equiv \frac 2{\hbar ^2\Omega ^2}
E^2=n(n+1),\,\,n\geq 0.
\]
Then this gives for the excitation spectrum
$E_n=\hbar \Omega \sqrt{\frac{n(n+1)}2}$ \cite{stri}.

\subsection{The correlation functions}

In this Subsection we calculate the two-point correlation function 
for the non-homogeneous gas as defined by Eq. (\ref{tag2}),
\begin{equation}
\Gamma (x_1,\tau _1;x_2,\tau _2)=\left\langle T_\tau
\hat \psi ^{\dagger}(x_1,\tau _1)\hat \psi (x_2,\tau _2)\right\rangle .
\label{opcf}
\end{equation}
The case of multi-point correlators will be studied in Section V.

We can express the correlator (\ref{opcf}) as the ratio of two functional
integrals \cite{pop}, \cite{pop1}:
\begin{equation}
\Gamma (x_1,\tau _1;x_2,\tau _2)=Z^{-1}\int e^{S[\psi ,\bar \psi ]}\bar \psi
(x_1,\tau _1)\psi (x_2,\tau _2){\cal D}\psi {\cal D}\bar \psi ,  
\label{fidcf}
\end{equation}
where the action $S[\psi ,\bar \psi ]$ is (\ref{act}) and $Z$ is the
partition function Eq. (\ref{part}).

We are interested in the long distance asymptotics of the correlators when
$\mid x_1-x_2\mid \gg l_c$.\ The main contribution to the asymptotics is
given by the low-lying excitations. By integrating out the fields $\psi _e$,
$\bar \psi _e$ included up to quadratic terms in Eq. (\ref{fidcf}), we find
that the leading term of the asymptotics is equal to
\begin{equation}
\Gamma (x_1,\tau _1;x_2,\tau _2)\simeq
\frac{\int e^{S_{eff}[\psi _o,\bar \psi _o]}
\bar \psi _o(x_1,\tau _1)\psi _o(x_2,\tau _2){\cal D}\psi _o{\cal D}
\bar\psi_o}
{\int e^{S_{eff}[\psi _o,\bar \psi _o]}{\cal D}\psi _o{\cal D}\bar \psi _o},
 \label{fcf}
\end{equation}
where $S_{eff}[\psi _o,\bar \psi _o]$ is the effective action (\ref{seff}).
We may rewrite this expression in terms of the density-phase variables
\begin{equation}
\Gamma (x_1, \tau_1; x_2, \tau_2)\simeq
\frac{\int \exp{\Bigl(S_{eff}[\rho ,\varphi ]-
i\varphi (x_1,\tau _1)+i\varphi (x_2,\tau _2)+
\frac 12\ln \rho (x_1,\tau_1)+\frac 12\ln \rho (x_2,\tau _2)\Bigr)}
{\cal D}\rho {\cal D}\varphi }
{\int \exp{\bigl(S_{eff}[\rho,\varphi ]\bigr)}{\cal D}\rho {\cal D}\varphi },
 \label{fcfpd}
\end{equation}
where $S_{eff}[\rho ,\varphi ]$ is Eq. (\ref{sseff}).

At low enough temperatures we may change $\ln \rho (x_1,\tau _1),\ln \rho
(x_2,\tau _2)$ in (\ref{fcfpd}) to $\ln \rho _{TF}(x_1),\ln \rho _{TF}(x_2)$
with $\rho _{TF}$ given by Eq. (\ref{tf}), since in the Thomas-Fermi regime
density fluctuations are suppressed \cite{df}, \cite{kr}. The functional
integrals in (\ref{fcfpd}) can be evaluated by steepest descents
\cite{pop1}, and the correlation function
$\Gamma (x_1, \tau_1; x_2, \tau_2)$ can be expressed in the form
$$
\Gamma (x_1,\tau _1;x_2,\tau _2)\simeq \exp \Bigl(
 -S_{eff}[\rho _0,\varphi_0]+S_{eff}[\rho _1,\varphi _1]
$$
\begin{equation}
\qquad\qquad\qquad
- i\varphi _1(x_1,\tau _1)+i\varphi_1(x_2,\tau _2)
           +\frac 12\ln \rho _{TF}(x_1)+\frac 12\ln \rho_{TF}(x_2)
\Bigr) ,
\label{fcfu}
\end{equation}
where the term $S_{eff}[\rho _0,\varphi _0]$ is given by Eq. (\ref{fe1}),
and the fields $\rho _0,\varphi _0$ satisfy two stationary
Gross-Pitaevskii equations (\ref {gp}). By definition, the fields
$\rho_1, \varphi_1$ are determined by the extremum condition:
\begin{equation}
\delta \left( S_{eff}[\rho ,\varphi ]-i\varphi (x_1,\tau _1)+i\varphi
(x_2,\tau _2)+\frac 12\ln \rho _{TF}(x_1)+\frac 12\ln \rho _{TF}(x_2)\right)
=0.  \label{fve}
\end{equation}
This variational equation leads to another pair of the Gross-Pitaevskii type
equations. The first equation appears as a coefficient at the variation
$\delta\rho(x,\tau )$,
\begin{equation}
 i\partial _\tau \varphi +\frac{\hbar ^2}{2m}\left( \frac 1{\sqrt{\rho }}%
\partial _x^2\sqrt{\rho }-\left( \partial _x\varphi \right) ^2\right)
+\left( \Lambda -V(x)\right) -g\rho =0,
\label{fgpcf1}
\end{equation}
while the second one is the coefficient at the variation $\delta\varphi (x,\tau )$,
\begin{equation}
 -i\partial _\tau \rho +\frac{\hbar ^2}m\partial _x\left( \rho
\partial _x\varphi \right)  =i\delta (x-x_1)\delta (\tau -\tau _1)-
i\delta (x-x_2)\delta (\tau -\tau_2).
\label{fgpcf2}
\end{equation}
Note that $\rho _{TF}(x)$ is a fixed function, and it is not subjected 
to variation. Note also that the equation Eq. (\ref{fgpcf2}) is 
driven by $\delta$-functions but Eq. (\ref{fgpcf1}) is not -- 
corresponding to the suppression of density fluctuations.
After substitution of the solutions $\rho _1,\varphi _1$ of 
Eqs. (\ref{fgpcf1}), (\ref{fgpcf2}) into the effective action 
Eq. (\ref{sseff}), one obtains
\begin{equation}
S_{eff}[\rho _1,\varphi _1]=-\beta F_{nc}(\mu )-1+\frac g2\int_0^\beta d\tau
\int dx\rho _1^2.  \label{fe2}
\end{equation}

We can furthermore consistently assume that, away from the boundaries, the
density fluctuations are small so that $\rho _1(x,\tau )=\rho _{TF}(x)+\pi
_1(x,\tau )$, and $\pi _1$ is a slowly varying function. Consequently the
terms $\sqrt{\pi _1}\partial _x^2\sqrt{\pi _1}$ and
$\partial _x\pi _1 \partial _x\varphi _1$
are small and can be dropped, and Eqs.~(\ref{fgpcf1}), (\ref{fgpcf2}), when
linearized around the Thomas-Fermi solution, can be expressed in the form
\begin{eqnarray}
i\partial _\tau \varphi _1-g\pi _1-\frac{\hbar ^2}{2m}\left( \partial
_x\varphi _1\right) ^2 &=&0,  \label{gpc1} \\
-i\partial _\tau \pi _1+\frac{\hbar ^2}m\partial _x\left( \rho _{TF}\partial
_x\varphi _1\right) &=&i\delta (x-x_1)\delta (\tau -\tau _1)-i\delta
(x-x_2)\delta (\tau -\tau _2).  \label{gpc2}
\end{eqnarray}

By differentiating Eq. (\ref{gpc1}) with respect to $\tau $, substituting
the result into Eq. (\ref{gpc2}), and dropping the terms of order higher
than $g$ or $\hbar ^2$, we find that
\begin{equation}
\frac 1g\partial _\tau ^2\varphi _1+\frac{\hbar ^2}m\partial _x\left( \rho
_{TF}(x)\partial _x\varphi _1\right) =i\delta (x-x_1)\delta (\tau -\tau
_1)-i\delta (x-x_2)\delta (\tau -\tau _2).  \label{ecf}
\end{equation}
It is convenient to rewrite this equation in the form
\begin{equation}
\frac 1{\hbar ^2v^2}\partial _\tau ^2\varphi _1+\partial _x\left( \tilde \rho
_{TF}(x)\partial _x\varphi _1\right) =i\frac{mg}{\hbar ^2\Lambda }\Bigl\{
\delta (x-x_1)\delta (\tau -\tau _1)-\delta (x-x_2)\delta (\tau -\tau
_2)\Bigr\} ,  \label{ecff}
\end{equation}
where $v$ is the sound velocity at the center of the trap, Eq. (\ref{sv}),
 and $\tilde \rho _{TF}$ is defined in Eq. (\ref{tf}). The solution of
this equation depends on the coordinates $x_1,\tau _1,x_2,\tau _2$ of the
$\delta $-sources so that $\varphi _1(x,\tau )\equiv 
\varphi _1(x,\tau;x_1,\tau _1,x_2,\tau _2).$ Now, with the use of 
Eqs. (\ref{fe1}), (\ref{fe2}),
(\ref{gpc1}), and (\ref {gpc2}), one can calculate the 
following contribution into  the exponent in Eq. (\ref{fcfu}) which is
\begin{eqnarray*}
&&\ \ \ \ \ \ -S_{eff}[\rho _0,\varphi _0]+S_{eff}[\rho _1,\varphi _1] \\
\ &\simeq &\frac g2\int_0^\beta d\tau \int dx\left( \rho _1^2-\rho
_0^2\right) =\frac g2\int_0^\beta d\tau \int dx\left( \rho _1-\rho _0\right)
\left( \rho _1+\rho _0\right) \\
\ &\simeq &g\int_0^\beta d\tau \int dx\pi _1\rho _0=\int_0^\beta d\tau \int
dx\left( i\partial _\tau \varphi _1-\frac{\hbar ^2}{2m}\Bigl( \partial
_x\varphi _1\Bigr) ^2\right) \rho _0 \\
\ &=&-\frac{\hbar ^2}{2m}\int_0^\beta d\tau \int dx\rho _0\left( \partial
_x\varphi _1\right) ^2=\frac{\hbar ^2}{2m}\int_0^\beta d\tau \int dx\varphi
_1(x)\partial _x\left( \rho _0(x)\partial _x\varphi _1(x)\right) \\
\ &=&\frac i2\left( \varphi _1(x_1,\tau _1)-\varphi _1(x_2,\tau _2)\right) .
\end{eqnarray*}
By substituting this expression into Eq. (\ref{fcfu}), we obtain for the
correlation function
$$
\Gamma (x_1,\tau _1;x_2,\tau _2)\simeq \sqrt{\rho _{TF}(x_1)\rho _{TF}(x_2)}
\exp{\Bigl(-\frac i2\left( \varphi _1(x_1,\tau _1)-\varphi _1(x_2,\tau _2)
\right)\Bigr) },
$$
where $\rho _{TF}(x)$ is given by Eq. (\ref{tf}).

It is convenient to express the solution of Eqs. (\ref{ecf}), (\ref{ecff})
in terms of the solution $G(x,\tau ;x^{\prime },\tau ^{\prime })$ of
the related equation
\begin{equation}
\ \ \ \frac 1{\hbar ^2v^2}\partial _\tau ^2G(x,\tau ;x^{\prime },\tau
^{\prime })+\partial _x\left( \Bigl( 1-\frac{x^2}{R_c^2}\Bigr) \partial
_xG(x,\tau ;x^{\prime },\tau ^{\prime })\right) =\frac g{\hbar ^2v^2}\delta
(x-x^{\prime })\delta (\tau -\tau ^{\prime }).
\label{g}
\end{equation}
and we call this equation as the `$\delta$-function driven
Stringari equation'. This gives for the first order correlation
function at $x_1\neq x_2$:
\begin{eqnarray}
\Gamma (x_1,\tau _1;x_2,\tau _2) &\simeq &\sqrt{\rho _{TF}(x_1)\rho
_{TF}(x_2)}  \label{focf} \\
&&\ \times \exp{\Bigl(-\frac 12\left( G(x_1,\tau _1;x_2,\tau _2)+G(x_2,\tau
_2;x_1,\tau _1)\right) +\frac 12G(x_1,\tau _1;x_1,\tau _1)+\frac 12%
G(x_2,\tau _2;x_2,\tau _2)\Bigr)}.  \nonumber
\end{eqnarray}
The function $G(x_1,\tau _1;x_2,\tau _2)$ has the sense of the correlation
function of phases,
\begin{equation}
G(x_1,\tau _1;x_2,\tau _2)=\bigl\langle \hat \varphi (x_1,\tau _1)\hat \varphi
(x_2,\tau _2)\bigr\rangle .  
\label{phasecf}
\end{equation}
The substitution of this formula into (\ref{focf}) gives the well known
result \cite{pop}
\begin{equation}
\Gamma (x_1,\tau _1;x_2,\tau _2)\simeq \sqrt{\rho _{TF}(x_1)\rho _{TF}(x_2)}
\exp{\Bigl(-\frac 12\bigl\langle \left( 
    \hat \varphi (x_1,\tau _1)-\hat \varphi (x_2,\tau
_2)\right) ^2\bigr\rangle\Bigr) }.  
\label{p}
\end{equation}

Notice that the terms in the exponent of the formula (\ref{focf}) have
different senses. The terms $G(x_1,\tau _1;x_2,\tau _2)$, $G(x_2,\tau
_2;x_1,\tau _1)$ in it depend on the relative position of the coordinates
and thus define the long distance behaviour of the correlator. The terms 
$G(x_1,\tau _1;x_1,\tau _1),G(x_2,\tau _2;x_2,\tau _2)$ each depend on a
single set of coordinates and thus contribute to the amplitudes only. Thus
we may express the correlation function in the form
\begin{equation}
\Gamma (x_1,\tau _1;x_2,\tau _2)\simeq
\sqrt{\tilde \rho (x_1,\tau _1)\tilde\rho (x_2,\tau _2)}
\exp{\Bigl(-\frac 12\left( G(x_1,\tau _1;x_2,\tau _2)+G(x_2,\tau
_2;x_1,\tau _1)\right)\Bigr) },  \label{focf2}
\end{equation}
where $\tilde \rho (x_1,\tau _1),\tilde \rho (x_2,\tau _2)$ are the 
renormalized densities \cite{bogoli}. The solution 
$G(x_1,\tau _1;x_2,\tau _2)$ of Eq. (\ref{g}) is defined up to an 
imaginary constant which has the sense of a global phase, and, 
as follows from (\ref{focf}), it does not influence the phase 
fluctuations.

The results of Section III are thus such that the excitation spectrum is
given by the solution of the Stringari equation (\ref{str}), while the
correlation of phases $\bigl\langle \hat \varphi (x_1,\tau _1)\hat \varphi
(x_2,\tau _2)\bigr\rangle $ is determined by $G(x_1,\tau _1;x_2,\tau _2)$
which is the solution of the $\delta $-function driven Stringari equation 
(\ref{g}).

\section{Asymptotic behaviours of the correlation functions}

\subsection{Homogeneous gas}

Here we apply the method presented in the previous section, Section III, to
the calculation of the asymptotics of the correlation function for the
homogeneous gas. To describe the case when the trap is absent,
$V(x)\equiv 0$, we may equally send the radius of the condensate
$R_c$ to infinity. This way the Eq. (\ref{g}) will take the form
\begin{equation}
\frac 1{\hbar ^2v^2}\partial _\tau ^2G(x,\tau ;x^{\prime },\tau ^{\prime
})+\partial _x^2G(x,\tau ;x^{\prime },\tau ^{\prime })=\frac g{\hbar ^2v^2}
\delta (x-x^{\prime })\delta (\tau -\tau ^{\prime }).  \label{hg}
\end{equation}
The solution of this equation is the correlation function of phases. We
consider Eq. (\ref{hg}) in the ranges of arguments
$0\leq x\leq L$ and $0\leq \tau\leq \beta $. For
$|x-x^{\prime }|\leq L,\,|\tau -\tau ^{\prime }|\leq \beta$, and
$\beta^{-1}\equiv k_BT\gg (\hbar v)/L$, we obtain the solution
\begin{equation}
G(x,\tau ;x^{\prime },\tau ^{\prime })=\frac g{2\pi \hbar v}\ln \left\{
2\left| \sinh \frac \pi {\hbar \beta v}\left( \left| x-x^{\prime }\right|
+i\hbar v(\tau -\tau ^{\prime })\right) \right| \right\} 
-\frac g{\hbar^2v^2}\frac{|x-x^{\prime }|^2}{2\beta L}.
\label{cfph}
\end{equation}
In the limit $L\rightarrow \infty $ the second term in this expression
disappears, and the substitution of the remaining expression in Eq. (\ref
{focf}) gives the expected answer Eq. (\ref{tcor}) for the asymptotics with
the scaling exponent equal to
\begin{equation}
\theta =\frac{2\pi \hbar v}g.  \label{exp}
\end{equation}
Using $v=\sqrt{\Lambda/m}$ and $\rho =\Lambda /g$ for the sound
velocity and density, respectively, one obtains the expression Eq. (\ref{isv})
for the scaling exponent. For small coupling constants we may re-express the
result (\ref{exp}) in the form of Eq. (\ref{sgexp}).

These asymptotics for the correlation functions are valid for the ring-shaped
Bose-Einstein condensates, and the complete agreement between Eq. (\ref{tcor})
found by the exact Bethe Ansatz method and the results of the functional
integral methods in this case of homogeneous gas, confirms the validity of
the functional integral method itself as described in its actual details
above. And this should mean that the results achievable by the functional
integral method in the inhomogeneous case $V(x)\neq 0$ should be equally
good. The case $V(x)\neq 0$ cannot (so far) be treated by any Bethe Ansatz
method as already mentioned.

\subsection{Trapped gas}

Let us consider the non-homogeneous Stringari equation (\ref{g}) which
determines through its solution $G(x, \tau; x^{\prime }, \tau^{\prime})$
the long distance behaviour of the correlation functions:
\begin{equation}
\frac 1{\hbar ^2v^2}\partial _\tau ^2G(x,\tau ;x^{\prime },\tau ^{\prime
})+\partial _x\left( \Bigl( 1-\frac{x^2}{R_c^2}\Bigr) \partial _xG(x,\tau
;x^{\prime },\tau ^{\prime })\right) =\frac g{\hbar ^2v^2}\delta
(x-x^{\prime })\delta (\tau -\tau ^{\prime }).  \label{old}
\end{equation}
We are looking for the solutions which are finite, periodic in $\tau $, and
quadratically integrable at $|x|<R_c$. The substitution of the series
\begin{equation}
G(x,\tau ;x^{\prime },\tau ^{\prime })=\frac 1\beta \sum_\omega e^{i\omega
(\tau -\tau ^{\prime })}G_\omega (x,x^{\prime }),  \label{sub}
\end{equation}
where $\omega =\frac{2\pi n}\beta ,n=0,\pm 1,...,$ into Eq. (\ref{old}) leads
to an equation for the spectral density,
\begin{equation}
-\frac{\omega ^2}{\hbar ^2v^2}G_\omega (x,x^{\prime })
+\frac d{dx}\left(\Bigl(1-\frac{x^2}{R_c^2}\Bigr) 
\frac d{dx}G_\omega (x,x^{\prime })\right)=\frac g{\hbar^2 v^2}
\delta (x-x^{\prime }).
\label{tleg}
\end{equation}

The solution of this equation is expressed in terms of the Legendre
functions of the first and second kind, $P_\nu (x/R_c)$ and $Q_\nu (x/R_c)$,
which are linearly independent solutions of the Legendre
equation Eq. (\ref{leg1}) with $\nu $ taken to be
$$
\nu =-\frac 12+
\sqrt{\frac 14-\Bigl( \frac{R_c}{\hbar v}\Bigr) ^2\omega ^2}\,.
$$
We get:
\begin{equation}
G_\omega (x,x^{\prime })
=\frac{gR_c}{2\hbar^2 v^2}\,\epsilon (x-x^{\prime})
\Bigl\{ Q_\nu \bigl(\frac x{R_c}\bigr) P_\nu\bigl(\frac{x^{\prime }}{R_c}\bigr)
-Q_\nu \bigl(\frac{x^{\prime }}{R_c}\bigr) P_\nu\bigl(\frac x{R_c}\bigr)\Bigr\} ,
\label{fg}
\end{equation}
where $\epsilon (x-x^{\prime })$ is the sign function: $\epsilon (x)\equiv
sign(x).$ The validity of this statement can be checked by substitution of
the result (\ref{fg}) into Eq. (\ref{tleg}), and with the help of the equality
\cite{hob}
\[
Q_\nu ^{\prime }(y)P_\nu (y)-Q_\nu (y)P_\nu ^{\prime }(y)=(1-y^2)^{-1}.
\]

For a stationary, $\tau $-independent correlator we have
$G(x;x^{\prime })=\frac 1\beta G_0(x,x^{\prime }),$ where
\begin{eqnarray}
G_0(x,x^{\prime }) &=&\frac{gR_c}{2 \hbar^2 v^2}\,\epsilon (x-x^{\prime})
\Bigl\{ Q_0\bigl(\frac x{R_c}\bigr)-
        Q_0\bigl(\frac{x^{\prime }}{R_c}\bigr)\Bigr\}
\nonumber \\
\ &=&\frac{gR_c}{4\hbar ^2v^2}\ln \left[\frac{\Bigl( 1+\frac{\mid
x_1-x_2\mid }{2R_c}\Bigr) ^2-\frac{(x_1+x_2)^2}{4R_c^2}}{\left( 1-\frac{%
\mid x_1-x_2\mid }{2R_c}\right) ^2-\frac{(x_1+x_2)^2}{4R_c^2}}\right] ,
\label{fgst}
\end{eqnarray}
and we have used here the particular properties of the Legendre functions
$P_0(x)=1$ and $Q_0(x)=\frac 12\ln \frac{1+x}{1-x}.$ This result leads to the
following correlation function \cite{pet},\cite{bogoli},\cite{lux},\cite{bo}
\begin{equation}
\Gamma (x_1;x_2)\simeq \sqrt{\rho _{TF}(x_1)\rho _{TF}(x_2)}\exp \Biggl(
-\frac{g R_c}{4 \hbar^2v ^2} \ln\left[ \frac{\Bigl( 1+\frac{\mid
x_1-x_2\mid }{2R_c}\Bigr) ^2-\frac{(x_1+x_2)^2}{4R_c^2}}{\Bigl( 1-\frac{
\mid x_1-x_2\mid }{2R_c}\Bigr) ^2-\frac{(x_1+x_2)^2}{4R_c^2}}\right]\Biggr),
\label{ln}
\end{equation}
where $\theta $ is the scaling exponent as given in Eq. (\ref{isv}).

To study the behaviour of the $\tau $-dependent correlation functions we
must first notice that the Green's function of Eq. (\ref{tleg}) is defined
up to a solution of the homogeneous Legendre equation Eq. (\ref{leg1}). To
guarantee the convergence of the series (\ref{sub}), we may add such a term
to the expression (\ref{fg}), and obtain thereby for the Green's
function, with $|\omega |>0$,
$$
G_\omega (x,x^{\prime }) =
\frac{gR_c}{2 \hbar^2 v^2} \epsilon (x-x^{\prime })
\Bigl\{
Q_\nu \bigl(\frac x{R_c}\bigr) P_\nu \bigl(\frac{x^{\prime }}{R_c}\bigr)
-Q_\nu \bigl(\frac{x^{\prime }}{R_c}\bigr) P_\nu \bigl(\frac x{R_c}\bigr)
\Bigr\}
$$
\begin{equation}
\qquad\qquad\quad
- i \frac{gR_c}{2 \hbar^2 v^2}
\Bigl\{
\frac 2\pi Q_\nu\bigl(\frac x{R_c}\bigr)Q_\nu\bigl(\frac{x^{\prime }}{R_c}
\bigr)
+\frac \pi 2P_\nu \bigl(\frac{x^{\prime }}{R_c}\bigr)P_\nu
\bigl(\frac x{R_c}\bigr)
\Bigr\}.
\label{fgw}
\end{equation}
For $|\omega |\gg \hbar v/(2R_c)$, the asymptotics of the Legendre
functions are \cite{hob}:
\begin{eqnarray*}
P_\nu (\cos \theta ) &\simeq &\Bigl( \frac 2{\pi \nu \sin \theta }\Bigr)^{1/2}
\sin \left[ \bigl(\nu +\frac 12\bigr)\theta +\frac \pi 4\right] , \\
Q_\nu (\cos \theta ) &\simeq &\Bigl( \frac \pi {2\nu \sin \theta }\Bigr)^{1/2}
\cos \left[ \bigl(\nu +\frac 12\bigr)\theta +\frac \pi 4\right] ,
\end{eqnarray*}
where $\cos \theta \equiv x/R_c,\, 0<\theta <\pi $, and $|\arg \nu |< \pi/2$.
Substituting these expressions into (\ref{fgw}) we find the
behaviour of the Green's function for large $|\omega |$,
\begin{equation}
G_\omega (x,x^{\prime })\simeq -\frac g{2\hbar v|\omega |}\frac 1{\sqrt{\sin
\theta \sin \theta ^{\prime }}}\exp \Bigl( -\frac{R_c}{\hbar v}|\omega
|\,|\theta -\theta ^{\prime }|\Bigr) .  \label{asg}
\end{equation}

In the quasi homogeneous limit, when $\left| x_1-x_2\right| 
\gg \frac{x_1+x_2}2$
and the coordinates $x_1$ and $x_2$ are far away from the boundaries
defined by $R_c$, so that $\left|x_1-x_2\right|\ll R_c$,
the leading term of the Green's function Eq. (\ref{fgst}) is
\begin{equation}
G_0(x,x^{\prime })\simeq \frac \Lambda {2\hbar ^2v^2\rho _{TF}(S)}\left|
x-x^{\prime }\right| .  \label{cfo}
\end{equation}
Here $S$ is the center-of-mass coordinate $S=\frac{x_1+x_2}2,$ and $v$ is
the sound velocity Eq. (\ref{sv}). For the functions with $\omega \neq 0$
and at inverse temperatures larger than the lowest energy excitations,
$\beta \equiv (k_BT)^{-1}\gg \hbar v/(2R_c)$, we may use the
asymptotics Eq. (\ref{asg}) to obtain
\begin{equation}
G_\omega (x,x^{\prime })\simeq -\frac \Lambda {2\hbar v\rho _{TF}(S)}
\frac{\exp{\Bigl(-(\hbar v)^{-1 }\mid \omega \mid \mid x-x^{\prime }\mid 
\Bigr)}} {\mid \omega\mid }.
\label{goo}
\end{equation}
The substitution of Eqs. (\ref{cfo}) and (\ref{goo}) in Eq. (\ref{sub})
gives now
\begin{equation}
G(x,\tau ;x^{\prime },\tau ^{\prime })=\frac \Lambda {2\pi \hbar v\rho
_{TF}(S)}\ln \left\{ 2\left| \sinh \frac \pi {\hbar \beta v}\left( \left|
x-x^{\prime }\right| +i\hbar v(\tau -\tau ^{\prime })\right) \right|
\right\} .  \label{tcfph}
\end{equation}

The correlation function for the Bose fields is in this case
\begin{equation}
\Gamma (x_1,\tau _1;x_2,\tau _2)\simeq \frac{\sqrt{\rho _{TF}(x_1)\rho
_{TF}(x_2)}}{\left| \sinh \frac \pi {\hbar \beta v}\left( \left|
x_1-x_2\right| +i\hbar v(\tau _1-\tau _2)\right) \right| ^{1/\theta (S)}},
\label{fcfnh}
\end{equation}
where the scaling exponent $\theta (S)$ depends now on the center-of-mass
coordinate $S=\frac{x_1+x_2}2$:
\[
\theta (S)=\frac{2\pi \hbar \rho _{TF}(S)}{mv}.
\]
From the result (\ref{fcfnh}) it follows that the correlation function
decreases exponentially at large distances,
\begin{equation}
\ \Gamma (x_1,\tau _1;x_2,\tau _2)\simeq \sqrt{\rho _{TF}(x_1)\rho _{TF}(x_2)
}\exp \left( -\frac{\Lambda \left| x_1-x_2\right| }{2\beta \hbar ^2v^2\rho
_{TF}(S)}\right) ,
\label{ld}
\end{equation}
in which the correlation length depends on the center-of-mass coordinate and
is
\begin{equation}
\xi (S)\equiv \frac{2\hbar ^2\beta \rho _{TF}(S)}m=\frac{\hbar \beta v}\pi
\theta (S).  \label{tcl}
\end{equation}
At zero temperature we find correspondingly that
\begin{equation}
\Gamma (x_1,\tau _1;x_2,\tau _2)\simeq \frac{\sqrt{\rho _{TF}(x_1)\rho
_{TF}(x_2)}}{\Bigl| \left| x_1-x_2\right| +i\hbar v(\tau _1-\tau _2)\Bigr|
^{1/\theta (S)}}.
\label{ztcfnh}
\end{equation}
These expressions for the trapped gas are similar to those for the
homogeneous gas, Eqs. (\ref{tcor}) and (\ref{iztcf}), but the scaling
exponents now depend on the center-of-mass coordinate.
We thus find that our results for the trapped Bose gas precisely assume
the forms of the exactly known results in the limit of vanishing
trap potential.

There is also another way to study the asymptotic behaviour of the
$\tau $-dependent correlation function in the quasi-homogeneous limit
\cite{bogoli}.
We may suppose that the term $(2x/R_c^2)\partial _xG(x,\tau ;x^{\prime
},\tau ^{\prime })$ in Eq. (\ref{old}) can be neglected, and that the factor
$\bigl( v^2\tilde \rho _{TF}(x)\bigr) ^{-1}$ in the first term in Eq. (\ref
{old}) can be substituted by the inverse squared sound velocity at the
center of the trap. The corresponding equation for the correlation function
of the phases in the quasi homogeneous limit is
\begin{equation}
\frac 1{\hbar ^2v^2}\partial _\tau ^2G(x,\tau ;x^{\prime },\tau ^{\prime
})+\partial _x^2G(x,\tau ;x^{\prime },\tau ^{\prime })=\frac g{\hbar ^2v^2
\tilde \rho _{TF}(\frac{x+x^{\prime }}2)}\delta (x-x^{\prime })\delta (\tau
-\tau ^{\prime }).  \label{tdegf}
\end{equation}
It is easy to check that the solution of this equation is Eq. (\ref{tcfph}).

\section{Multi-particle correlation functions}

We consider finally the asymptotic behaviour of the multi-particle
temperature-dependent correlation functions of the Bose fields
$\hat \psi^{\dagger },\hat \psi $,
\begin{equation}
\Gamma _m(x_1,\tau _1;...;x_{2m},\tau _{2m})\equiv \langle T_\tau \hat \psi
^{\dagger }({\bf z}_1)...\hat \psi ^{\dagger }({\bf z}_m)\hat \psi ({\bf z}
_{m+1})...\hat \psi ({\bf z}_{2m})\rangle ,  \label{mcfd}
\end{equation}
where ${\bf z}=(x,\tau )$. We show in particular that at low temperatures
and for large relative distances $R_{ij}\equiv\mid x_i-x_j\mid $, the
$m$-particle correlation function can be expressed in terms of the 
two-point ones,
\begin{equation}
\Gamma _m(x_1,\tau _1;...;x_{2m},\tau _{2m})\simeq \prod_{1\leq i<j\leq
2m}\Gamma (x_1,\tau _1;x_2,\tau _2)^{-l_il_j},  \label{mcfr}
\end{equation}
where $l_i=1$ if $i=1,...,m$, $l_i=-1$ if $i=m+1,...,2m$, and
$\Gamma (x_i,\tau _i;x_j,\tau _j)
=\bigl\langle T_\tau\hat \psi ^{\dagger }(x_i,\tau _i)\hat \psi
(x_j,\tau _j)\bigr\rangle $
is the two-point correlation function discussed in
the previous sections. We suppose as well that the points $x_i$ are lying
inside the condensate and sufficiently far from its boundaries. The
factorisation of the asymptotic behaviour of the correlation function
Eq. (\ref{mcfr}) is similar to that in the non-trapped case \cite{popmp};
the lack of the translational invariance only introduces modifications 
to the two-point correlations.

We can express Eq. (\ref{mcfd}) as the ratio of two functional integrals
over the space of complex-valued functions (c.f. Section III):
\begin{equation}
\ \Gamma _m(x_1,\tau _1;...;x_{2m},\tau _{2m})=Z^{-1}
\int e^{S [\psi,\bar\psi]}
\bar \psi(x_1,\tau _1)...\bar \psi (x_m,\tau _m)\psi (x_{m+1},\tau _{m+1})...
\psi(x_{2m},\tau _{2m}){\cal D}\psi {\cal D}\bar \psi ,
\label{mfi}
\end{equation}
where $Z$ is the partition function Eq. (\ref{part}). In order to estimate
$\Gamma _m$, Eq. (\ref{mfi}), we proceed as in Section III. The leading
term of the asymptotics is as follows:
\begin{equation}
\Gamma _m(x_1,\tau _1;...;x_{2m},\tau _{2m})\simeq \frac{
\int e^{S_{eff}[\psi_o, \bar\psi_o]}
\bar \psi _o({\bf z}_1)...\bar \psi _o({\bf z}_m)\psi _o({\bf z}
_{m+1})...\psi _o({\bf z}_{2m}){\cal D}\psi _o{\cal D}
                    \bar \psi _o}{\int e^{S_{eff}}{\cal D}\psi_o{\cal D}
              \bar\psi _o},
\label{mcf}
\end{equation}
where $S_{eff}$ is the effective action Eq. (\ref{seff}). Repeating the
arguments of Section III, we can rewrite the expression for the 
correlation function in the form
\begin{equation}
 \Gamma_m (x_1, \tau_1; ...; x_{2m}, \tau_{2m})\simeq
\frac{
\int \exp{\Bigl(
S_{eff}[\rho ,\varphi ]-i\!\sum\limits_{k=1}^m\varphi ({\bf z}_k)+
i\!\!\!\sum\limits_{k=m+1}^{2m}\varphi ({\bf z}_k)
+\frac 12\sum\limits_{k=1}^{2m}\ln \rho _{TF}({\bf z}_k)\Bigr)}
                           {\cal D}\rho {\cal D}\varphi }
{\int \exp{\bigl(S_{eff}[\rho ,\varphi ]\bigr)}{\cal D}\rho {\cal D}\varphi }.
\label{mcfpd}
\end{equation}

At low enough temperatures these remaining functional integrals can be
evaluated by steepest descents, and the correlation function takes the form
$$
 \Gamma _m(x_1,\tau _1;...;x_{2m},\tau _{2m})
\simeq \exp\biggl(-S_{eff}[\rho _o,\varphi _o]+S_{eff}[\rho _1,\varphi_1]
$$
\begin{equation}
-i\sum_{k=1}^m\varphi _1({\bf z}_k)+i\!\!\!
                           \sum_{k=m+1}^{2m}\varphi _1({\bf z}_k)
+\frac 12\sum_{k=1}^{2m}\ln \rho _{TF}({\bf z}_k)\biggr),
\label{mcfu}
\end{equation}
where the fields $\rho _0,\varphi _0$ satisfy two stationary
Gross-Pitaevskii Eqs. (\ref{gp}). By definition, the fields
$\rho _1,\varphi _1$ are determined by the extremum condition
\begin{equation}
\delta \Bigl( S_{eff}[\rho ,\varphi ]-i\sum_{k=1}^m\varphi ({\bf z}%
_k)+i\!\!\!\sum_{k=m+1}^{2m}\varphi ({\bf z}_k)+\frac 12\sum_{k=1}^{2m}
\ln \rho_{TF}({\bf z}_k)\Bigr) =0,
\label{mve}
\end{equation}
and this variational equation also leads to a pair of the
Gross-Pitaevskii type equations
\begin{eqnarray*}
&i&\partial _\tau \varphi +\frac{\hbar ^2}{2m}\left( \frac 1{\sqrt{\rho }}
\partial _x^2\sqrt{\rho }-\left( \partial _x\varphi \right) ^2\right)
+\left( \Lambda -V(x)\right) -g\rho  =0 \\
-&i&\partial _\tau \rho +\frac{\hbar ^2}m\partial _x\left( \rho \partial
_x\varphi \right)  =i\sum_{k=1}^m\delta (x-x_k)\delta (\tau -\tau
_k)-i\!\!\!\sum_{k=m+1}^{2m}\delta (x-x_k)\delta (\tau -\tau _k).
\end{eqnarray*}

In the approximation considered in Subsection III.B (see Eqs.
(\ref{gpc1}), (\ref{gpc2})), these equations lead us to the
following equation:
\begin{equation}
\frac 1{\hbar ^2v^2}\partial _\tau ^2\varphi +\partial _x\left( \tilde \rho
_{TF}(x)\partial _x\varphi \right) =i\frac{mg}{\hbar ^2\Lambda }\Bigl\{
\sum_{k=1}^m\delta (x-x_k)\delta (\tau -\tau _k)-\!\!\!\sum_{k=m+1}^{2m}
\delta (x-x_k)\delta (\tau -\tau _k)\Bigr\} .
\label{mecf}
\end{equation}
One finds now that
$$
-S_{eff}[\rho _0,\varphi _0]+S_{eff}[\rho _1,\varphi _1]=\frac i2\Bigl(
\sum_{k=1}^m\varphi _1({\bf z}_k)-
\!\!\!\sum_{k=m+1}^{2m}\varphi _1({\bf z}_k)\Bigr) .
$$
Substitution of this expression in Eq. (\ref{mcfu}) leads to the correlation
function
\begin{equation}
\Gamma _m(x_1,...,x_{2m})\simeq \prod_{k=1}^{2m}\sqrt{\rho _{TF}(x_k)}
\exp{\biggl(-\frac i2\Bigl( \sum_{k=1}^m\varphi _1({\bf z}_k)-
\!\!\!\sum_{k=m+1}^{2m}\varphi _1({\bf z}_k)\Bigr)\biggr) },
\label{cfa}
\end{equation}
where the function $\varphi _1({\bf z})\equiv \varphi _1({\bf z;z}_1,...,%
{\bf z}_{2m})$ satisfies Eq. (\ref{mecf}). We can then express the solution
of Eq. (\ref{mecf}) in terms of the function $G({\bf z},{\bf z}^{\prime }),$
which is the solution of Eq. (\ref{g}), and this yields
$$
-\frac i2\Bigl( \sum_{k=1}^m\varphi _1({\bf z}_k)-
\!\!\!\sum_{k=m+1}^{2m}\varphi_1({\bf z}_k)\Bigr)
         =-\frac 12\sum_{j=1}^m\sum_{k=m+1}^{2m}\left( G({\bf z}_j,{\bf z}_k)
                                   +G({\bf z}_k,{\bf z}_j)\right)
$$
\begin{equation}
 +\frac 12\sum_{j=1}^{m-1}\sum_{k=j+1}^m
\left( G({\bf z}_j,{\bf z}_k)+G({\bf z}_k,{\bf z}_j)\right)
+\frac 12\sum_{j=m+1}^{2m-1}\sum_{k=j+1}^{2m}
\left( G({\bf z}_j,{\bf z}_k)+G({\bf z}_k,{\bf z}_j)\right)
+\frac 12\sum_{j=1}^{2m}G({\bf z}_j,{\bf z}_j).
\label{mfg}
\end{equation}
The first order correlation function at ${\bf z}_1\neq {\bf z}_2$ is then
equal to that of Eq. (\ref{focf}). Substitution of this expression in 
Eq. (\ref{mfg}) gives the already indicated result Eq. (\ref{mcfr}):
\begin{eqnarray*}
\Gamma _m({\bf z}_1,...,{\bf z}_{2m})=\prod_{1\leq i<j\leq 2m}\Gamma
^{-l_il_j}({\bf z}_i,{\bf z}_j).
\end{eqnarray*}

\section{Conclusions}

It is plain that the functional integral methods used here to calculate
correlation functions of 1D bosons are very powerful. They extend the
methods introduced by Popov \cite{pop},\cite{pop1} to the case with a
confining potential that breaks the translational symmetry of the system. A
key new ingredient introduced here is the Green's function $G$ that satisfies
the $\delta $-function driven Stringari equation Eq. (\ref{g}). This Green's
function was shown to be precisely the correlation of phases. We thus obtain
the results quoted in \cite{pet},\cite{bogoli} in a very natural way, 
for the Green's function could then be used to obtain the two-point 
correlation function. Note how density fluctuations were suppressed 
by the step at which $\delta $-function driving terms were deliberately 
neglected from the Eq. (\ref{gpc1}). It is rather remarkable that for 
a vanishing trap potential the two-point correlation function found by the 
approximated functional-integral method coincides {\it precisely} 
with the exact Bethe Ansatz result, especially when one takes into 
account the changed boundary conditions.

The 'universal' feature in the two-point correlation functions obtained with
or without a trap potential is the scaling exponent $\theta $ that governs
the decay of correlations for increasing separation of the two points. At
non-zero temperatures $\theta $ is proportional to the correlation length
and at zero temperature it becomes a critical exponent in the algebraic
decay of correlations. We express this scaling exponent in terms of the
observed quantities, sound velocity $v$ and the density $\rho $ of the quasi
condensate, which means we need not calculate the $T$-matrix as, {\it e.g.},
in \cite{khaw}. It also means that the expression we find for $\theta $ is
valid for all coupling strengths: the two physical quantities involved do
depend on coupling strength so that $\theta =2$ in the exact Girardeau-Tonks
limit $g\to \infty $, and $\theta >2$ for any finite $g$. In the presence of
a confining potential the decay of correlations is formally exactly the same,
but the scaling exponent will now depend on the center-of-mass coordinate,
$$
\theta (S)=\frac{2\pi \hbar \rho _{TF}(S)}{mv}
$$
with $S=\frac{x_1+x_2}{2}$. As we expect the $S$-dependence of $\theta $ to
be small, it should approach two in the Girardeau-Tonks limit also 
in this case. In the weak-interaction Gross-Pitaevskii limit $\theta$ 
can also be expressed in the form
$$
\theta (S)=2\pi \hbar \sqrt{\frac{\rho _{TF}^2(S)}{mg\rho _{TF}(0)}}.
$$
As all quantities that enter these expressions for $\theta $ are themselves
observables, a measurement of the decay of correlations in a trapped Bose
system would provide a sensitive test of the theoretical framework based on
repulsive $\delta $-function interactions between the particles. With a
sensitive enough measurement one should also be able to detect the effects
of confinement, {\it i.e.}, the dependence on the center-of-mass coordinate
$S$ in the decay of the correlations.

As experiments on trapped Bose condensates have become increasingly
more sophisticated, a need for better understanding of also many-particle
correlations has become evident also. So far little has been known about the
$2m$-point correlation functions for $m>1$. One of the important results 
of this paper is thus the demonstration that all $2m$-point correlations 
in trapped Bose systems can be expressed asymptotically entirely in 
terms of the two-point correlations. Note that, based on the results 
given here, and the exact results we have reported earlier on a related Bose 
system \cite{bogol}, we can also propose that the asymptotic form of 
the density-density correlator for trapped Bose systems as
\begin{eqnarray*}
&&\ \left\langle T_\tau \hat \psi ^{\dagger }(x_1,\tau _{1)}\hat \psi
(x_1,\tau _1)\hat \psi ^{\dagger }(x_2,\tau _2)\hat \psi (x_2,\tau
_2)\right\rangle -\rho _{TF}(x_1)\rho _{TF}(x_2) \\
\  &\sim &A\left( \frac 1{w^2}+\frac 1{\bar w^2}\right) +\frac B{|w|^{\theta
(S)}}\cos \left[ 2\pi (x_1-x_2)\rho _{TF}(S)\right]
\end{eqnarray*}
with $w=\sinh \left\{ (\pi |x_1-x_2|/\hbar \beta v)+i\hbar v(\tau _1-\tau
_2)\right\} $. Here $A$ and $B$ are real constants. At large separations of
points this density-density correlator decays as $Ae^{-\zeta |x_1-x_2|}$
with the correlation length $\zeta =(2\pi )^{-1}\hbar \beta v$. Notice that
the scaling exponent $\theta $ appears here in the oscillating term which is
not the leading term asymptotically. One would nevertheless expect that 
the oscillating behaviour in the asymptotic density-density correlations
should be experimentally observable.

Finally we would like to emphasize that the methods applied here are not
restricted to 1D but can equally well be applied in 2D and 3D 
\cite{bogoli},\cite{tmf}. These methods are not restricted to the 
calculation of correlation functions, and can be used to calculate 
all of the thermodynamic properties of interacting Bose systems, and
can be extended beyond the Thomas-Fermi as well
as the mean-field approximations.

\section*{Acknowledgement}
N.M.B. would like to thank the Royal Society of London and
the Department of Physics, University of Jyv\"askyl\"a
for support. 
N.M.B. and C.M. acknowledge that this paper was partially supported by
RFBR (Project No. 01-01-01045).
This work has also been supported by the Academy of Finland
under the Center of Excellence Program (Project No. 44875).
We acknowledge gratefully the invaluable help of
J.B. Parkinson with finalizing this paper.

\section*{Appendix A}

Here we shall give a formal derivation of the effective action
$S_{eff}[\psi_o,\bar\psi_o]$, Eq. (\ref{seff}). We
use the field-theoretical approach of
loop expansion \cite{zub}. Following the general prescription of Section
III, we split the initial fields $\psi ,\bar\psi $ into $\psi =\psi_o
+\psi_e$, $\bar\psi =\bar\psi_o+\bar\psi_e$ so that the action
$S[\psi ,\bar\psi ]$, Eq. (\ref{act}), is divided into three terms:
$$
S=S_{cond}+S_{free}+S_{int},
\eqno(A1)
$$
where $S_{cond}$ is the action of the condensed particles in the so-called
tree-level approximation \cite{ram}
$$
S_{cond}[\psi_o,\bar \psi_o]\equiv \int_0^\beta d\tau \int dx\left\{ \bar
\psi_o(x,\tau )\hat K_{+}\psi_o(x,\tau )-\frac g2\bar\psi_o(x{\bf ,}
\tau )\bar\psi_o(x{\bf ,}\tau )
             \psi_o(x{\bf ,}\tau )\psi_o(x{\bf ,}\tau)\right\}.
\eqno(A2)
$$
The action of the highly excited thermal fluctuations of the non-condensed
particles is given by
$$
S_{free}[\psi_e,\bar \psi_e]\equiv \frac 12\int_0^\beta d\tau \int dx
(\bar \psi_e,\psi_e) \hat G^{-1}
\left(
\begin{array}{c}
\psi_e \cr
\bar \psi_e
\end{array}
\right) ,
\eqno(A3)
$$
and $S_{int}$ is the part of the action that describes the interaction of
condensed particles with the thermal ones,
$$
S_{int}[\psi_o,\bar\psi_o,\psi_e,\bar\psi_e]\equiv \int_0^\beta
d\tau \int dx \Bigl\{\bar\psi_e(x,\tau )
\left[ \hat{K}_{+}-g \bar\psi_o\psi_o\right]
\psi_o(x,\tau )+ \psi _e(x,\tau )
\left[ \hat{K}_{-}-g\bar\psi_o\psi_o\right] \bar\psi_o(x,\tau )
\Bigr\}.
\eqno(A4)
$$
In Eqs. (A2)--(A4) we have defined the differential operators,
$$
\hat{K}_{\pm } 
\equiv \pm \frac \partial {\partial \tau }-{\cal H}, \qquad\qquad
{\cal H} =-\frac{\hbar ^2}{2m}\frac{\partial ^2}{\partial x^2}-\mu +V(x),
$$
and the matrix operator
$$
\hat{ G}^{-1} \equiv \hat{G}_0^{-1}-\hat{\Sigma },
\eqno(A5)
$$
where
$$
\hat{G}_0^{-1} \equiv
\left(
\begin{array}{cc}
\hat{K}_{+} & 0 \\
                  0 & \hat{K}_{-}
\end{array}
\right) ,
\qquad\hat{\Sigma }\equiv \hat{\Sigma }(\psi_o,\bar \psi_o)=g
\left(
\begin{array}{cc}
2\bar\psi_o \psi _o & \psi_o^2 \\
(\bar \psi_o)^2 & 2\bar\psi_o\psi_o
\end{array}
\right).
$$
The approximation used here for $S_{int}$, Eq. (A4), implies that the
non-condensed excitations do not interact with each other.

For our purposes it is enough to consider the functional integral in leading
order as given by the method of stationary phase (steepest descents).
Therefore we choose $\bar \psi_o,\psi_o$ as the stationary points of
$S_{cond}$ (A2). They are governed by equations of the
Gross--Pitaevskii type,
$$
\begin{array}{r}
\displaystyle{
\left( \frac \partial {\partial \tau }+\frac{\hbar ^2}{2m}\frac{\partial ^2}
{\partial x^2}+\mu -V(x)\right) \psi_o-g (\bar\psi_o \psi_o) \psi_o =0,}
 \\ [0.4cm]
\displaystyle{
\left(-\frac \partial {\partial \tau }+\frac{\hbar ^2}{2m}\frac{\partial ^2}
{\partial x^2}+\mu -V(x)\right) 
                     \bar \psi_o-g (\bar\psi_o\psi_o)\bar \psi_o =0.}\\
\end{array}
\eqno(A6)
$$
As soon as $\psi_o,\bar \psi_o$ are the solutions of these equations,
$S_{int}$ drops out of Eq. (A1), and the dynamics of $\psi_e,\bar
\psi_e$ is just given to lowest order by $S_{free}$,  Eq. (A3).
The latter depends non-trivially on $\bar \psi_o,\psi_o$ through
$\hat{\Sigma }$ in $\hat{G}^{-1}$, Eq. (A5). We denote a
$\tau $-independent solution of the Gross--Pitaevskii Eqs. (A6),
written in the Thomas--Fermi approximation as
$$
\bar\psi_o\psi _o=\rho _{TF}(x;\mu )\equiv \frac 1g\left(\mu
-V(x)\right) \Theta\left( \mu -V(x)\right) ,
\eqno(A7)
$$
where one should notice the dependence of the Thomas-Fermi profile on the
`bare' chemical potential $\mu$. The integration over $\psi _e,\bar \psi _e$ 
in Eq. (\ref{effact}) is now
Gaussian, and gives a one-loop corrected effective action \cite{pop} in
terms of the condensate variables $\psi_o,\bar \psi_o$ only,
$$
S_{eff}[\psi _o,\bar \psi_o]\equiv S_{cond}[\psi_o,\bar \psi_o]-
\frac 12 {\rm ln} \det (\hat{G}^{-1}).
\eqno(A8)
$$
Here $\hat{G}^{-1}$ is the matrix operator  (A5).

To assign a meaning to the resulting expression for the effective action
(A8), we have to regularize the determinant $\det (\hat{G}^{-1})$.
Our $\hat{G}^{-1}$ is already written in the form of a $2\times
2$ matrix Dyson equation (A5), where the entries of the matrix
$\hat{\Sigma }(\psi _o,\bar \psi _o)$ play the role of the normal and
anomalous self-energy parts. The Dyson equation defines the matrix
Green function $\hat{G}\equiv \hat{G}(\psi _o,\bar \psi _o)$
({\it i.e.}, the propagator matrix) of the fields $\psi_e,\bar \psi_e$.
The matrix $\hat{G}$ is given by the formal inverse
$$
\hat{G}=\left( \hat{G}_0^{-1}-\hat{\Sigma }\right) ^{-1}.
\eqno(A9)
$$
The matrix operator $\hat{G}^{-1}$ (A5) can be formally
diagonalized by means of the famous $(u,v)$-transformation of Bogoliubov
\cite{nnb},\cite{df}. Compatibility of the corresponding equations for the
unknown fields $u,v$ defines a quasi-classical spectrum of the elementary
excitations \cite{str}.

For our purposes it is appropriate to represent $\hat{G}^{-1}$ in the form
$$
\hat{G}^{-1}=\hat{G}_0^{-1}-\hat{\Sigma }\equiv \hat{{\cal G}}^{-1}-
\Bigl( \hat{\Sigma }-2g\rho _{TF}(x;\mu )\hat I\Bigr) ,
\eqno(A10)
$$
where $\hat I$ is a $2\times 2$ unit matrix, and $\hat{{\cal G}}^{-1}$
is defined as
$$
\hat{{\cal G}}^{-1}\equiv \left(
\begin{array}{cc}
\hat{K}_{+}-2g\rho _{TF}(x;\mu ) & 0 \\
0 & \hat{K}_{-}-2g\rho _{TF}(x;\mu )
\end{array}
\right) \equiv \left(
\begin{array}{cc}
{\cal K}_{+} & 0 \\
0 & {\cal K_{-}} \\
\end{array}
\right) .
\eqno(A11)
$$
Here $\rho_{TF}(x; \mu )$ is the solution (A7) of Eq. (A6), 
and the expression (A10) implies that we have simply added and 
subtracted $2g\rho_{TF}(x;\mu )$ in the diagonal of the matrix 
operator $\hat{G}^{-1}$. A formal inverse of $\hat{{\cal G}}^{-1}$ 
can be found from the equation which defines the
corresponding Green's functions ${\cal G}_{\pm }$:
$$
\left(
\begin{array}{cc}
{\cal K}_{+} & 0 \cr
0 & {\cal K_{-}}
\end{array}
\right) \left(
\begin{array}{cc}
{\cal G}_{+} & 0 \cr
0 & {\cal G}_{-}
\end{array}
\right) =\delta (x-x^{\prime })\delta (\tau -\tau ^{\prime })\hat{I}\,.
$$

Applying the relation ${\rm ln} \det ={\rm tr\, ln}$, which is
valid for both matrices and operators, we obtain
$$
-\frac 12{\rm ln} \det \hat{G}^{-1}=-\frac 12{\rm tr\, ln}
\left( \hat I-\hat{{\cal G}}(\hat{\Sigma }
-2g\rho _{TF}(x;\mu )\hat I)\right) -\frac 12{\rm ln}
\det \left(
\begin{array}{cc}
{\cal K}_{+} & 0 \\
0 & {\cal K_{-}}
\end{array}
\right) .
\eqno(A12)
$$
The first term on the r.h.s. is free of divergences and now we have to
assign a meaning to the infinite-order determinant on the r.h.s. of (A12).
The operators ${\cal K}_{\pm }$ may be expressed in the form
$$
{\cal K}_{\pm }\equiv \pm \frac \partial {\partial \tau }+
\frac{\hbar ^2}{2m}\frac{\partial ^2}{\partial x^2}+|V(x)-\mu |.
\eqno(A13)
$$
Notice how minus two times 
$\left(\mu -V(x)\right)\Theta\left(\mu -V(x)\right)$
in ${\cal K}_{+},{\cal K}_{-}$ introduces the absolute value
$|V(x)-\mu |$
into these differential operators. Let us denote the eigenvalues of
${\cal K}_{\pm }$ as $\pm i\omega -\lambda _n$, where $\omega $ is a
bosonic Matsubara frequency, and $\lambda _n$ is an eigenenergy of
the operator $-(\hbar ^2/2m)\partial ^2/\partial x^2-|V(x)-\mu |$. The
regular part of the
logarithm of the determinant in Eq. (A12) has the sense of the free
energy $\tilde{F}_{nc}$ of an ideal gas of thermal particles
\[
\tilde{F}_{nc}(\mu )\equiv \frac 1{2\beta }{\rm ln} \det \left(
\begin{array}{cc}
{\cal K}_{+} & 0 \\
0 & {\cal K_{-}}
\end{array}
\right) =\frac 1\beta \sum_n{\rm ln} \left( 2\sinh \frac{\beta \lambda_n}2
\right),
\]
where the regularized values of the determinants of the differential
operators ${\cal K}_{\pm}$ can be obtained, for instance, by means of a
zeta-regularization
procedure \cite{ram}. Then, up to a  first order in $g$, we find that
$$
-\frac 12 {\rm ln} \det \hat{G}^{-1} =-\beta \tilde{F}_{nc}(\mu)
+g\int_0^\beta d\tau \int dx\left( {\cal G}_{-}(x,\tau ;x,\tau )+
{\cal G}_{+}(x,\tau ;x,\tau )\right) (\bar\psi_o\psi_o-\rho_{TF}(x;\mu ))
$$
$$
\equiv -\beta F_{nc}(\mu )+2g\int_0^\beta d\tau \int dx\rho_{nc}(x)
\bar\psi_o\psi_o.
 \eqno(A14)
$$
Here $F_{nc}$ is the free energy of the non-ideal gas of thermal particles,
and the last term in (A14) describes the interaction of the thermal
particles with the condensate. The density of the non-condensed particles is
$\rho_{nc}(x)\equiv -{\cal G}_{\pm }(x,\tau ;x,\tau )$, and it depends,
in fact, on
the spatial variable only because of the translational invariance in $\tau$.
For very low temperatures and far from the boundaries of the condensate we
can use $\rho_{nc}(x)\simeq \rho_{nc}(0)$ because ${\cal G}_{\pm }(x,\tau
;x,\tau )$ is nearly a constant over most parts of the condensate \cite{nar}.

We now finally obtain the effective action
$$
S_{eff}[\psi_o,\bar \psi _o] =-\beta F_{nc}(\mu )
$$
$$
 +\int_0^\beta d\tau \int dx\left\{ \bar \psi _o(x,\tau )\left(
\frac{\hbar ^2}{2m}\frac{\partial ^2}{\partial x^2}+\Lambda -V(x)\right)
\psi_o(x,\tau )-\frac g2\bar \psi _o(x{\bf ,}\tau )\bar \psi_o(x{\bf ,}\tau
)\psi_o(x{\bf ,}\tau )\psi_o(x{\bf ,}\tau )\right\} ,
\qquad\qquad \eqno (A15)
$$
where $\Lambda =\mu -2g\rho _{nc}(0)$ is a renormalized chemical potential.
We consider the $S_{eff}$ (A15) as an effective one-loop action
which involves the thermal corrections above the {\it classical} level
(see Eqs. (A6)). Note that this derivation of the effective action does
not depend on the dimensionality of the system, and is thus valid in
two and three space dimensions as well.

\section*{Appendix B}

We derive here the scaling exponents Eqs. (\ref{tg}) and (\ref{sgexp}) for
infinite and small values of the coupling constant $g$, respectively, from
the {\it universal} result Eq. (\ref{isv}) based on the Bethe Ansatz
equations Eq. (\ref{cbethe}). It is possible to solve these equations
exactly in the limiting cases $g\rightarrow 0$ and $g\rightarrow \infty$.

Let us discuss first the weak-interaction limit $g\rightarrow 0$. In order
to study this free boson limit we have to rescale the parameters $\lambda_j$
in Eq. (\ref{cbethe}):$\,\,\lambda _j=(\tilde g/L)^{1/2}\mu _j$, and then
take $g\rightarrow 0.$ The Bethe equations for the ground state will now
take the form
$$
\mu _j=\sum_{k=1,k\neq j}^N\frac 1{\mu _j-\mu _k}\,;\,\,j=1,...,N.
$$
These equations are equations for the zeros of the Hermite polynomials.
By applying the equality
\[
\sum_{j=1}^N\sum_{k\neq j}\frac{x_j}{x_j-x_k}=\frac 12N(N-1),
\]
it becomes easy to calculate the ground-state energy of the weakly
interacting gas
$$
E_N=\frac gL\sum_{j=1}^N\mu _j^2=\frac g{2L}N(N-1).
$$
In the thermodynamic limit $N,L\rightarrow \infty $ with the density $\rho
=\frac NL$ fixed, the free energy ${\cal E}$ of the weakly interacting
Bose gas is
$$
{\cal E}=\lim_{N\rightarrow \infty }\frac{E_N}L=\frac g2\rho^2
$$
so that for small coupling constants the sound velocity is
$$
v=\sqrt{\frac{g\rho }m}.
$$
The substitution of this result in Eq. (\ref{isv}) gives the result (\ref
{sgexp}).

The case of impenetrable (hard core) bosons, also referred to as the
free-fermion limit or the Girardeau-Tonks gas, corresponds to an infinite
value of the coupling constant $g=\infty$. The Bethe equations Eq. (\ref
{cbethe}) now become
$$
e^{i\lambda_jL}=(-1)^{N+1}.
$$
These equations describe noninteracting particles and are easily solved,
$$
\lambda_j=\frac{2\pi }L\left( j-\frac{N+1}2\right) ,\,\,j=1,...,N,
\eqno(B1)
$$
and these solutions fill in the interval $\lambda_1\leq \lambda_j\leq
\lambda_N$ , where the Fermi radius $\lambda_F\equiv \lambda_N$
corresponds to the solution with $j=N$ ($-\lambda_F\equiv \lambda_1$):
$$
\lambda_F=\frac \pi L(N-1).
$$
By using the solutions of (B1), the ground state energy of the
Girardeau-Tonks gas is easily calculated to be
$$
E_N=\frac{\pi ^2\hbar ^2}{6mL^2}N(N^2-1).
$$
In the thermodynamic limit the free energy ${\cal E}$ of the Girardeau-Tonks
gas is
$$
{\cal E}=\frac{\pi ^2\hbar ^2}{6m}\rho ^3,
$$
and the sound velocity at zero temperature is
$$
v=\frac{\hbar \pi }m\rho .
$$
Substitution of this result in Eq. (\ref{isv}) gives the exponent of Eq.
(\ref{tg}).


\end{document}